\documentclass[pdflatex,sn-mathphys-num]{sn-jnl}
\usepackage{graphicx}
\usepackage{color}
\usepackage{amsmath}
\usepackage{amssymb}
\usepackage{natmove}
\usepackage{natbib}
\usepackage{hyperref}
\usepackage{bm}
\usepackage{comment}
\usepackage[normalem]{ulem}

\begin{document}

\title[Article Title]{Collapse of Jahn-Teller Phonons in La$_{1-x}$Sr$_{x}$MnO$_3$ with Weak Magnetoresistance}

\author*[1]{\fnm{Tyler C.} \sur{Sterling}}\email{ty.sterling@colorado.edu}
\author[2]{\fnm{Andrei T.} \sur{Savici}}
\author[3]{\fnm{Ryoichi} \sur{Kajimoto}}
\author[4,5]{\fnm{Kazuhiko} \sur{Ikeuchi}}
\author[6]{\fnm{Nazir} \sur{Khan}}
\author[6]{\fnm{Frank} \sur{Weber}}
\author*[1,7]{\fnm{Dmitry} \sur{Reznik}}\email{dmitry.reznik@colorado.edu}

\affil[1]{\orgdiv{Department of Physics}, \orgname{University of Colorado Boulder}, \orgaddress{ \city{Boulder}, \state{Colorado}, \postcode{80309},  \country{USA}}}

\affil[2]{\orgdiv{Neutron Scattering Division}, \orgname{Oak Ridge National Laboratory}, \orgaddress{ \city{Oak Ridge}, \state{Tennessee}, \postcode{37830},  \country{USA}}}

\affil[3]{\orgdiv{Materials and Life Science Division, J-PARC Center}, \orgname{Japan Atomic Energy Agency}, \orgaddress{ \city{Tokai}, \state{Ibaraki}, \postcode{319-1195},  \country{Japan}}}

\affil[4]{\orgdiv{Institute of Materials Structure Science}, \orgname{High Energy Accelerator Research Organization (KEK)}, \orgaddress{ \city{Tokai}, \state{Ibaraki}, \postcode{319-1106},  \country{Japan}}}

\affil[5]{\orgdiv{Neutron Science and Technology Center}, \orgname{Comprehensive Research Organization for Science and Society}, \orgaddress{ \city{Tokai}, \state{Ibaraki}, \postcode{319-1106},  \country{Japan}}}

\affil[6]{\orgdiv{Institute for Quantum Materials and Technologies}, \orgname{Karlsruhe Institute of Technology}, \orgaddress{ \postcode{D-76131}, \city{Karlsruhe}, \country{Germany}}}

\affil[7]{\orgdiv{Center for Experiments on Quantum Materials}, \orgname{University of Colorado Boulder}, \orgaddress{ \city{Boulder}, \state{Colorado}, \postcode{80309},  \country{USA}}}

\abstract{
Perovskite manganites are quantum materials exhibiting competing interactions inducing colossal magnetoresistance (CMR). The prevailing theory of CMR highlights the essential role of electron-phonon coupling (EPC), but mounting evidence suggests the underlying mechanism is more complicated. Here, we investigate phonons and spin–phonon coupling in ferromagnetic CMR manganites La$_{1-x}$Sr$_x$MnO$_3$ ($x$=0.2,0.3) with relatively small CMR associated with melting of the magnetic order above room temperature. High-resolution neutron scattering experiments combined with density functional theory (DFT) show that the low-temperature ferromagnetic phase is conventional: neutron scattering from phonons agrees with DFT predictions and magnons follow sinusoidal dispersions. Fluctuating magnetic moments and low-energy phonons remain conventional in the high temperature paramagnetic phase, indicating the Mn and La/Sr sublattices are not strongly perturbed by melting of ferromagnetism. In contrast, the Jahn–Teller–active optical oxygen vibrations collapse entirely above the Curie temperature, despite low CMR in these compositions, with some of the lost spectral weight reappearing as quasielastic scattering. We attribute this highly anomalous behavior to giant EPC in the charge and/or orbital channel. It drives cooperative diffusive motion of quasistatic carrier-trapping oxygen sublattice distortions once ferromagnetism disappears. We hypothesize the magnitude of magnetoresistance correlates with the rate of diffusion rather than with the strength of Jahn–Teller EPC.
}


\maketitle

\section{Introduction}

The perovskite manganites, R$_{1-x}$A$_{x}$MnO$_3$ (where R is a rare-earth and A an alkaline-earth metal) \cite{jonker1950ferromagnetic,van1950electrical}, are quantum materials exhibiting competing interactions that induce colossal magnetoresistance (CMR) \cite{von1993giant,jin1994thousandfold}. CMR is associated with coupled metal–insulator and ferromagnetic–paramagnetic transitions \cite{mizokawa2000orbital}. These materials serve as a model system for studying the interplay of spin, charge, orbital, and lattice degrees of freedom \cite{tokura2006critical,salamon2001physics,edwards2002ferromagnetism} and have potential applications in spintronics \cite{bibes2007oxide}. Early theories emphasized the double exchange mechanism \cite{zener1951interaction,anderson1955considerations,kubo1972quantum}, but later work highlighted the essential role of electron-phonon coupling (EPC) of Jahn-Teller type \cite{millis1995double,millis1996cooperative,millis1996dynamic} and stipulated that the magnitude of CMR, often inversely correlated with the Curie temperature $T_C$, scales directly with the strength of this EPC \cite{millis1996cooperative,millis1996dynamic,tokura2006critical,salamon2001physics}. However, mounting evidence suggests that the underlying mechanism is more complicated \cite{weber2013large,maschek2016polaronic,maschek2018polaronic}.  

\begin{figure*}
    \includegraphics[width=1.0\linewidth]{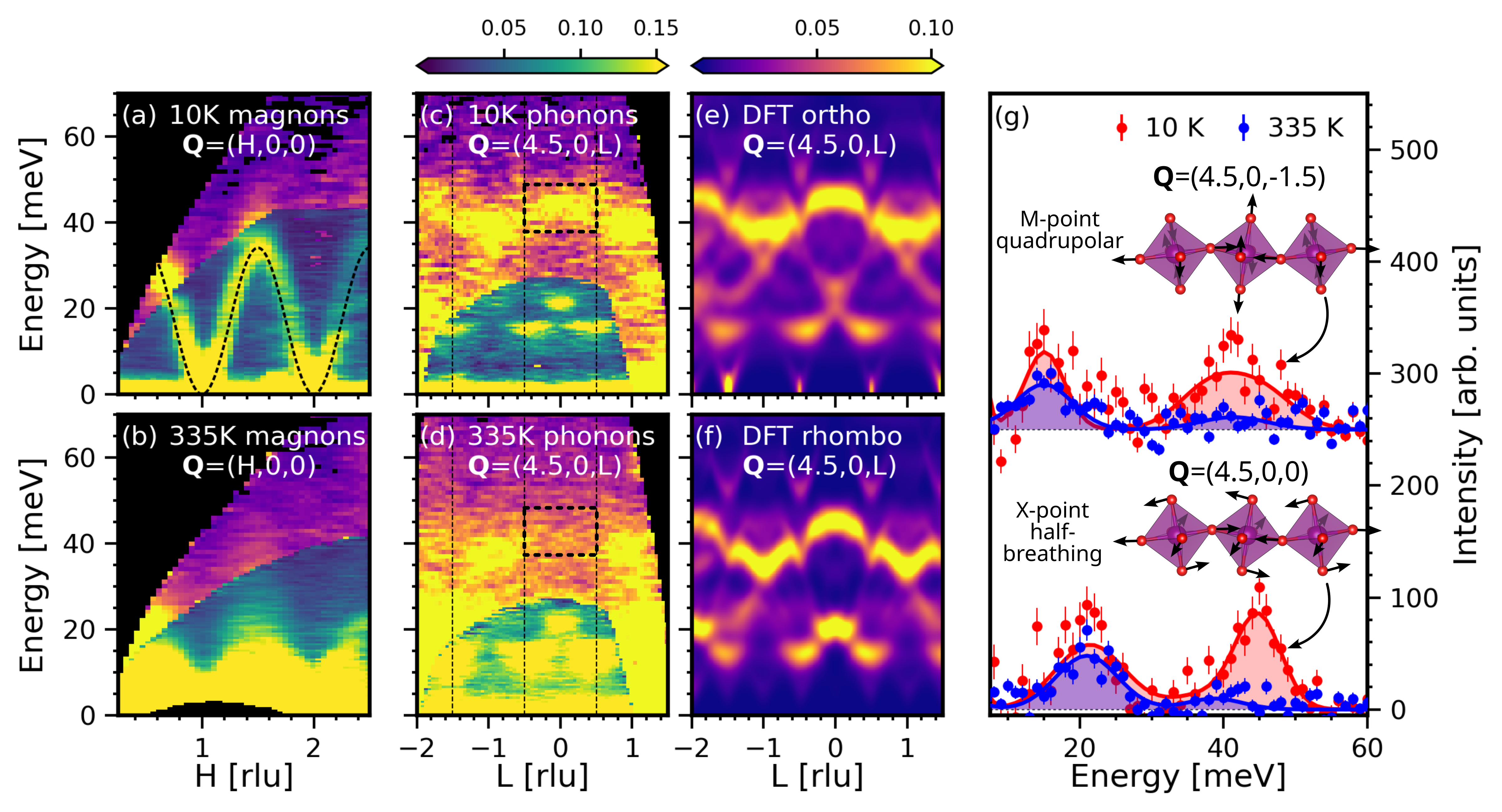}
    \caption{\textbf{Phonon anomalies in La$_{0.8}$Sr$_{0.2}$MnO$_3$.} Inelastic time-of-flight neutron scattering from the spin (a-b) and lattice degrees of freedom (c-d) in La$_{0.8}$Sr$_{0.2}$MnO$_3$ at 10 K (a,c) and 335 K (b,d). In (a-d), the upper/lower colormap is from $E_i=120/54$ meV. The dashed line in (a) is from a first-nearest neighbor Heisenberg model calculation using linear spin wave theory. (e-f) depict inelastic neutron scattering from phonons calculated from DFT. Line cuts through the $X$-point (in the cubic basis) half-breathing mode and the $M$-point quadrupolar mode, at $\approx 45$ meV at $\bm{Q}=(4.5,0,0)$ and $\approx 42.5$ meV at $\bm{Q}=(4.5,0,-1.5)$ respectively in (c), are compared at 10 K and 335 K in (g). The eigenvector diagrams depict the half-breathing and quadrupolar bond-stretching modes whose intensities are indicated the by the arrows in (g)}
    \label{fig:main_fig}
\end{figure*}

Earlier work reported anomalous behavior of the Jahn-Teller (JT) active phonons in large CMR compound La$_{0.7}$Ca$_{0.3}$MnO$_3$ where significant EPC effects are expected \cite{zhang2001jahn}. Surprisingly, studies on lower CMR compounds La$_{1-x}$Sr$_{x}$MnO$_3$ have also found evidence of strong EPC: the ferromagnetic transition strongly affects acoustic phonons and gives rise to quasielastic diffuse scattering in the high temperature paramagnetic phase \cite{weber2013large,maschek2016polaronic,maschek2018polaronic}. However, effect of EPC on JT active optical phonons implicated in CMR is not well understood. In this work, we employed neutron scattering to examine the phonons in La$_{0.7}$Sr$_{0.3}$MnO$_3$ and La$_{0.8}$Sr$_{0.2}$MnO$_3$ which have relatively high Curie temperatures of $T_C\approx350$ K and $T_C\approx305$ K and small magnetoresistance of about $10$ fold and $100$ fold, respectively; large CMR materials like La$_{0.7}$Ca$_{0.3}$MnO$_3$ have $T_C\approx250$ K and $1000$ fold CMR magnitude \cite{jiang2007extreme,mahesh1995giant,mccormack1994very,ju1994giant}.

Our inelastic neutron scattering measurements, supported by detailed density functional theory (DFT) calculations, revealed that an entire JT active bond-stretching optical phonon branch lost intensity and effectively collapsed away from the zone center above $T_C$ (Fig. \ref{fig:main_fig}c–d), despite only weak CMR and otherwise conventional magnetism in both La$_{0.7}$Sr$_{0.3}$MnO$_3$ and La$_{0.8}$Sr$_{0.2}$MnO$_3$. The missing inelastic weight appears to transfer to quasielastic scattering indicating diffusion of structural distortions. DFT calculations of lattice dynamics (Fig. \ref{fig:main_fig}e-f) agreed well with the low temperature experimental data and, combined with multizone analysis of neutron scattering intensities \cite{parshall2014phonon,reznik2020automating}, identified JT active character of the anomalous phonons (Fig. \ref{fig:main_fig}c-d,g). DFT rules out the orthorhombic-rhombohedral phase transition and twinning, both known to occur in these compounds, as the origin of the phonon collapse pointing instead at giant EPC of JT type. Our experiments and spin-wave calculations reveal negligible spin-phonon coupling originating from magnon-phonon hybridization and minimal magnon broadening (Fig. \ref{fig:main_fig}a) implying negligible spin-orbit and exchange contributions to the spin-phonon interaction \cite{woods2001magnon,furukawa1999magnon,sjostrand2022magnon}. The presence of strong, non-perturbative damping of the JT modes in Sr doped manganites with CMR as low as $10$ fold and Ca doped manganites with CMR of $1000$ fold suggests that CMR magnitude does not scale with the strength of JT EPC. Instead, we hypothesize that the diffusion of charge-trapping lattice distortions determines the magnitude of CMR: high CMR compounds host static or slow moving distortions, whereas low CMR compounds show comparatively rapid diffusion.

\begin{figure}
    \centering
    \includegraphics[width=0.5\linewidth]{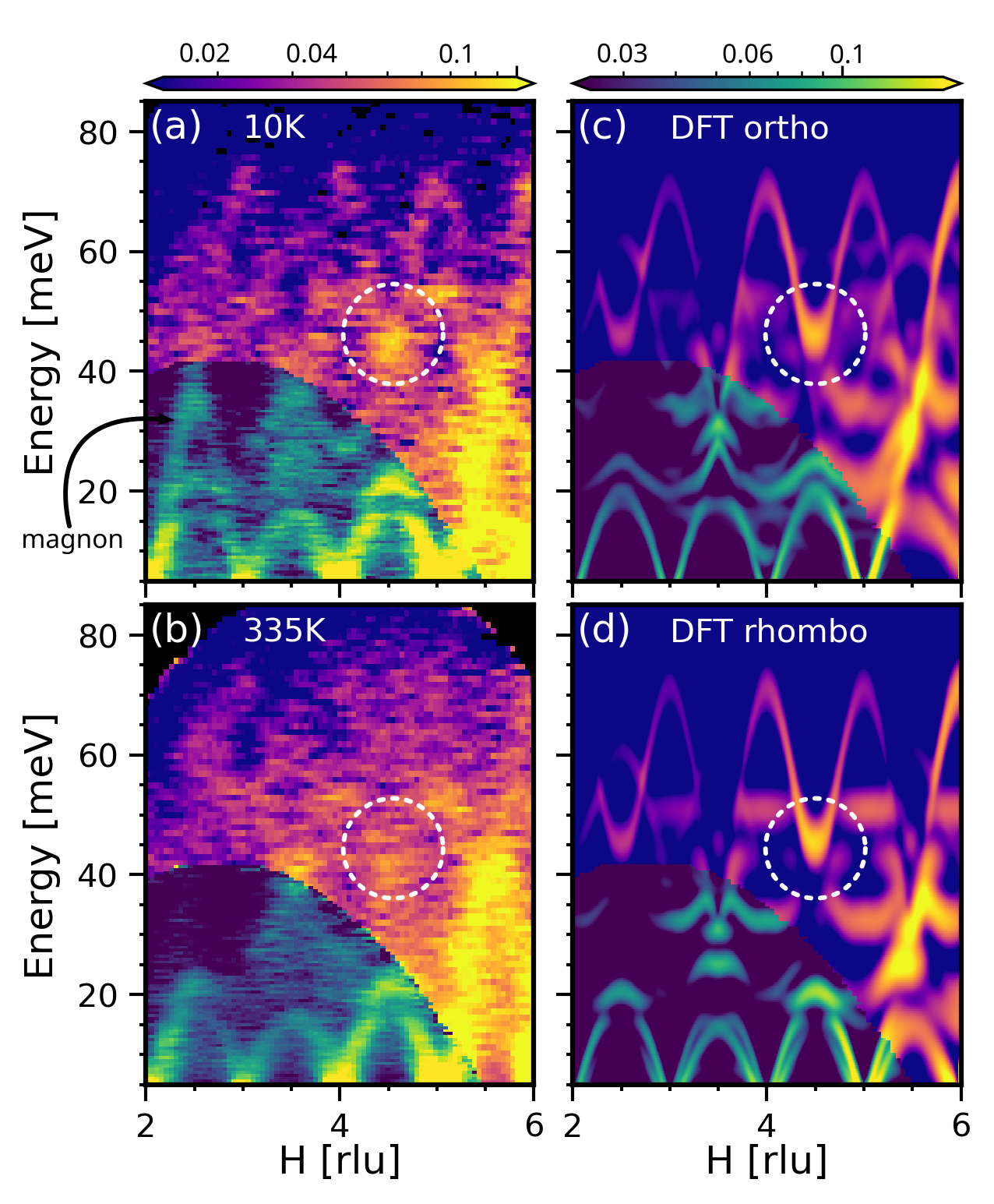}
    \caption{\textbf{Comparison of INS and DFT calculations.} Comparison between TOF neutron scattering measurements from La$_{0.8}$Sr$_{0.2}$MnO$_3$ at 10 K (a) and 335 K (b) and DFT calculations in the orthorhombic (c) and rhombohedral (d) phases along $\bm{Q}=(H,0,0)$ at 0 K. The upper/lower colormaps correspond to $E_i=120/54$ meV. The $E_i=120$ meV DFT calculations were broadened in energy using a Gaussian resolution with FWHM=5 meV; the $E_i=54$ meV calculations used FWHM=2.5 meV. The strong feature in the $2\leq H\leq 3$ zone at 10 K is a magnon which becomes quasielastic at 335 K. The white circle indicates the location of a strong phonon peak expected from DFT but appearing only at 10 K in the experiments. The data are divided by the Bose factor.}
    \label{fig:compare_ins}
\end{figure}

\begin{figure*}
    \centering
    \includegraphics[width=1.0\linewidth]{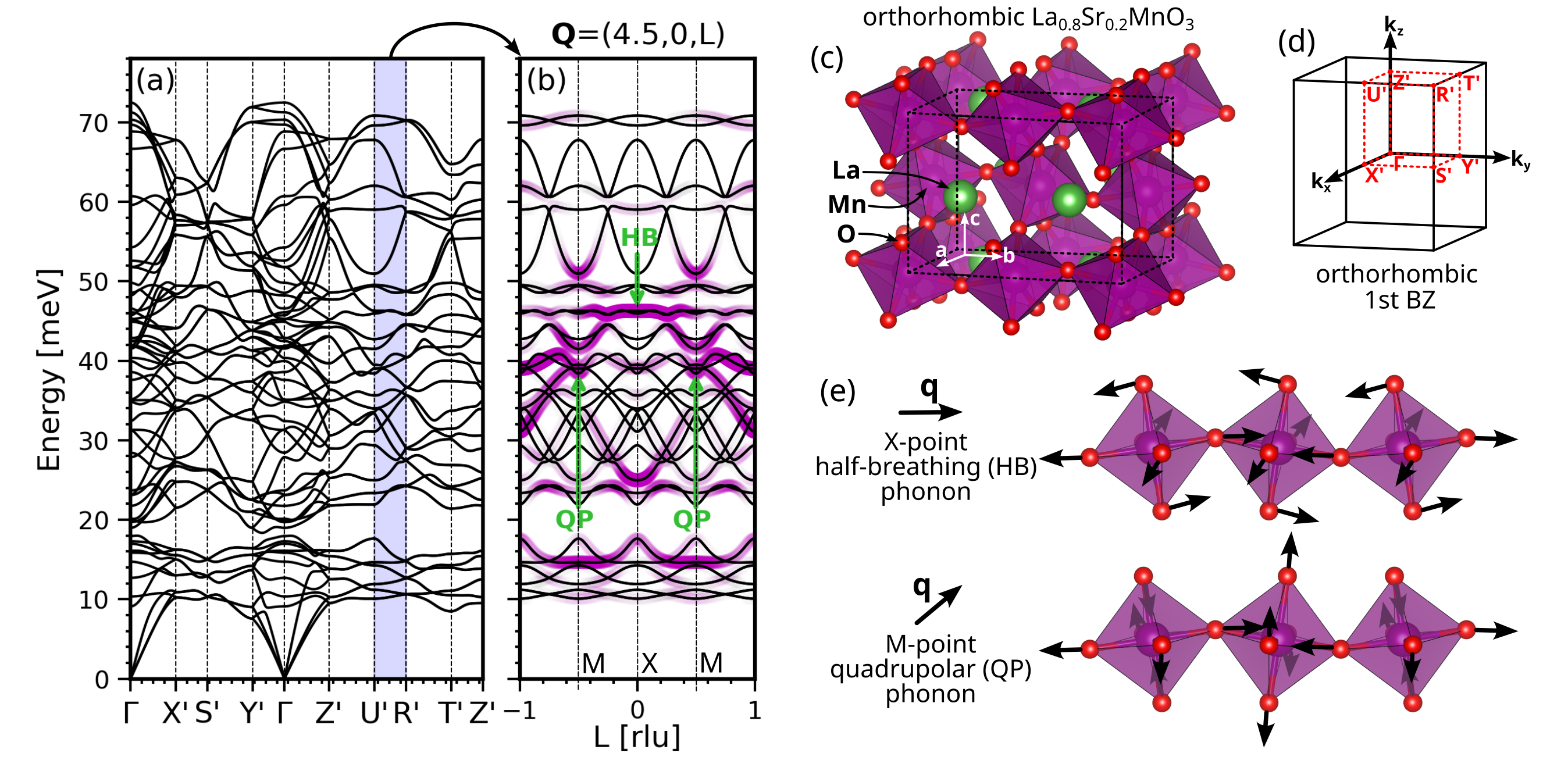}
    \caption{\textbf{Phonon dispersion calculations.} (a) Phonon dispersions in the orthorhombic phase of La$_{0.8}$Sr$_{0.2}$MnO$_3$. The primed labels are $\bm{q}$-points in the orthorhombic basis, while the unprimed $X$ and $M$ points in (b) correspond to the cubic basis. The orthorhombic unit cell is shown in (c) and the corresponding Brillouin zone for the same orientation is shown in (d). Panel (b) depicts the calculated inelastic scattering intensity (magenta shading) superposed with the phonon dispersions along the $\bm{Q}=(4.5,0,L)$ direction in the cubic basis (see Fig. \ref{fig:main_fig}e). The orientation in (b) has $L$ parallel to the orthorhombic $\bm{b}$ axis. This is the same as the orthorhombic $U'$-$R'$ direction as indicated by the blue shading in (a). In the orthorhombic basis in (b), $\bm{Q}=(4.5,2L,4.5)$. The eigenvector diagrams in (e) show the $X$-point half-breathing (HB) and $M$-point quadrupolar (QP) bond-stretching modes discussed in the text. The arrows in (b) indicate the location of the HB and QP modes in the dispersions. Only Mn-O octahedral chains along the $\bm{a}+\bm{c}$ direction are shown; the La atoms do not participate in the displacement, so are omitted.}
    \label{fig:phonon_dispersions}
\end{figure*}

\section{Results}

\subsection{Electronic structure and phonons from DFT}

Experimentally, below $\approx 120$ K La$_{0.8}$Sr$_{0.2}$MnO$_3$ is orthorhombic ($Pnma$); above $\approx 120$ K, it is rhombohedral ($R\bar3c$) \cite{paul1998influence,dabrowski1999structure,kamenev1997influence,asamitsu1996magnetostructural,tokura1994giant,urushibara1995insulator}. To gain insight into the anomalous phonon in Figure \ref{fig:main_fig} (c-d,g) and to rule out both structural phase transitions and twinning as the origin of the anomaly, we calculated the lattice dynamics and simulated inelastic neutron scattering intensity from La$_{0.8}$Sr$_{0.2}$MnO$_3$ using DFT (Figs. \ref{fig:main_fig}e-f and \ref{fig:compare_ins}c-d). We assume the results for La$_{0.7}$Sr$_{0.3}$MnO$_3$ to be similar, so we only calculate $x=0.2$ doping. 

The DFT approximation used in this work correctly predicts the electronic structure to be half-metallic with completely polarized Mn $t_{2g}$ core and $e_g$ conduction electrons in the cubic, orthorhombic, and rhombohedral phases. Moreover, the effect of charge doping on the electronic structure is to, more-or-less, rigidly shift the Fermi level down. The lower symmetry of the distorted unit cells lifts the degeneracy of some of the bands, but the electronic structures are qualitatively the same for all of the cubic, orthorhombic, and rhombohedral cells. See the Supplementary Information for more details \cite{suppinfo}.

We modeled the lattice dynamics in both the orthorhombic and rhombohedral phases and calculated intensities for all pseudo cubic-equivalent orientations in each phase to simulate different domains in a twinned crystal like those used for single-crystal neutron scattering. The orthorhombic unit cell is shown in Figure \ref{fig:phonon_dispersions} (c); the rhombohedral unit cell and the conversion from rhombohedral/orthorhombic to pseudocubic notation is discussed in the Supplementary Information \cite{suppinfo}. 

Neutron scattering intensity calculated using the DFT in both phases is in good agreement with the experimental data at 10 K (cf. Figs. \ref{fig:main_fig}c,e-f and \ref{fig:compare_ins}a,c-d). The energy-momentum slices calculated in the orthorhombic phase (Figs. \ref{fig:main_fig} (e) and \ref{fig:compare_ins} (c)) and in the rhombohedral phase (Figs. \ref{fig:main_fig} (f) and \ref{fig:compare_ins} (d)) correspond to the pseudocubic $(h,0,0)$ direction. In the high-temperature rhombohedral phase the phonon spectra along the pseudocubic crystal axes are equivalent and we show one cut. However, in the low-temperature orthorhombic phase, the phonon spectra along the pseudocubic axes are inequivalent and a direct comparison to the experimental data is not possible. Still, the spectra along each axis are separately in good agreement with the experimental data. Since it is impossible to distinguish the spectra from different domains in a twinned crystal, we average the orthorhombic calculations in Figures \ref{fig:main_fig} (e) and \ref{fig:compare_ins} (c) over all equivalent directions to compare to experimental data (see the Supplementary Information \cite{suppinfo} for a more detailed discussion). 

The eigenvectors in the rhombohedral phase have been studied in detail before \cite{reznik2005bond}, so we focus on the orthorhombic phase here. There are 20 atoms in the orthorhombic unit cell, resulting in 60 modes in the calculated dispersions in the orthorhombic phase (Fig. \ref{fig:phonon_dispersions}a). To relate these modes to the pseudocubic cell and identify the relevant anomalies revealed in the experiments, we plot the calculated neutron scattering intensities in the pseudocubic $\bm{Q}=(4.5,0,0)$ zone on top of the dispersions in Figure \ref{fig:phonon_dispersions} (b), enabling us to identify the relevant branches in the calculation. The eigenvectors of the anomalous $X$ and $M$ point branches in the orthorhombic phase, labeled in Figure \ref{fig:phonon_dispersions} (b), are depicted in Figures \ref{fig:phonon_dispersions} (e) and \ref{fig:main_fig} (g). We identify them as in-plane bond-stretching modes. Due to the underlying octahedral distortions, the normal modes are not pure bond-stretching modes: they mix with other branches, but are still similar to bond-stretching modes. At the $X$ point, the mode at $\approx 45$ meV is the half-breathing mode; at the $M$ point, the mode at $\approx 42.5$ is the quadrupolar mode (Fig. \ref{fig:phonon_dispersions}). The relevant eigenvectors in the rhombohedral phase have been studied before using shell model calculations \cite{reznik2005bond}; the analogous eigenvectors in the orthorhombic and rhombohedral phases have similar bond-stretching character.

The orthorhombic phase is predicted to be stable (i.e. no imaginary modes, cf. Fig. \ref{fig:phonon_dispersions}a). The calculated phonons in the rhombohedral phase showed an instability corresponding to octahedral rotations towards the orthorhombic phase which was expected since the rhombohedral phase is unstable at low temperature. Only the rotational mode is unstable and it is non-degenerate and well separated from the bond-stretching modes; in other words, the instability does not affect the relevant bond-stretching modes and we do not discuss it further here.

\subsection{Magnons in La$_{0.8}$Sr$_{0.2}$MnO$_3$}

\begin{figure*}
    \centering
    \includegraphics[width=1.0\linewidth]{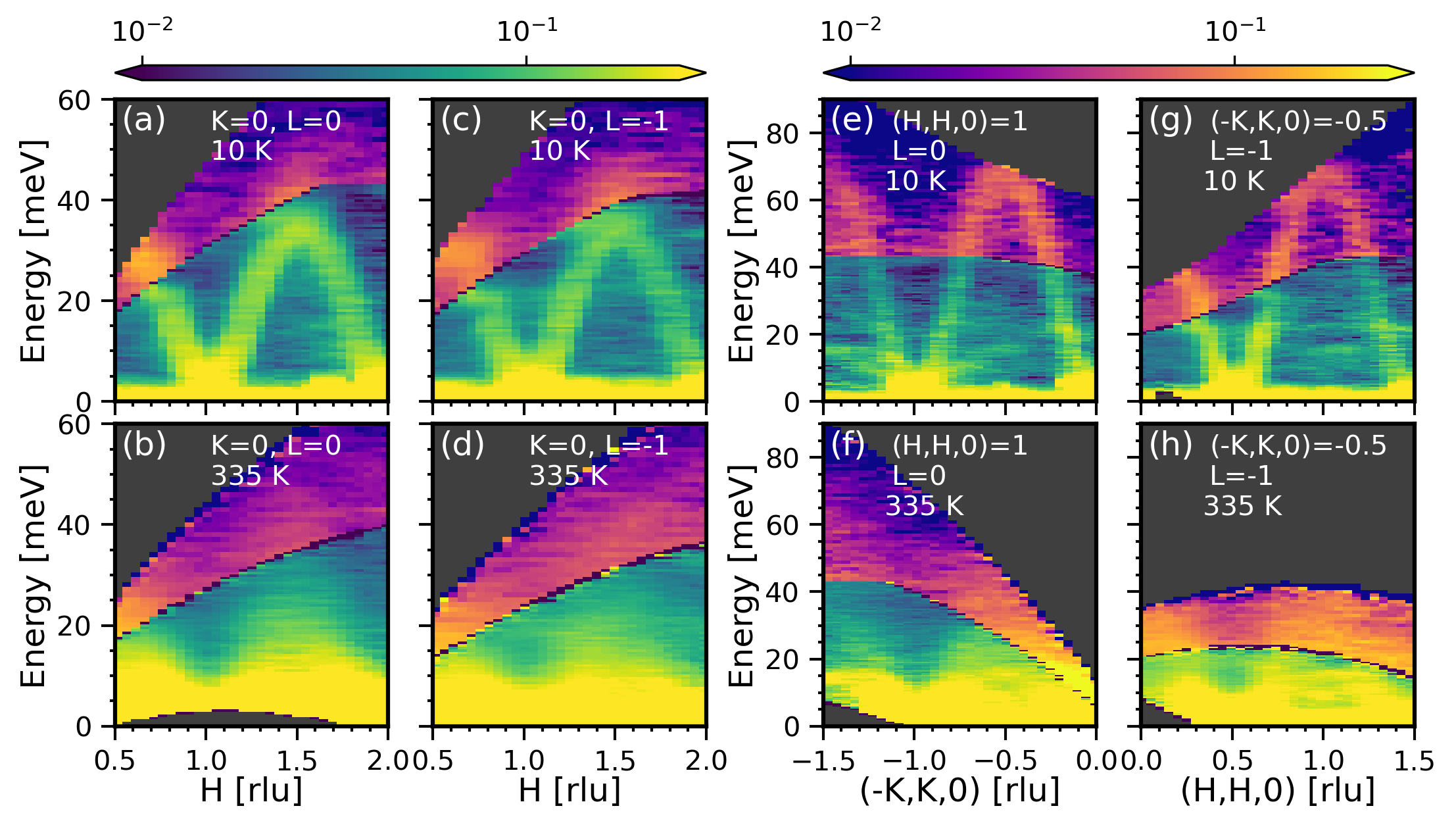}
    \caption{\textbf{Magnon dispersions in La$_{0.8}$Sr$_{0.2}$MnO$_3$ from time-of-flight neutron scattering.} Magnetic excitations in La$_{0.8}$Sr$_{0.2}$MnO$_3$ at 10 K (a,c,e,g) and at 335 K (b,d,f,h) measured by inelastic neutron scattering. The upper (lower) colormap is from $E_i=120$ (54) meV}
    \label{fig:magnons_vs_temperature}
\end{figure*}

Inelastic time-of-flight neutron scattering from La$_{0.8}$Sr$_{0.2}$MnO$_3$ reveals that the low-temperature spin dynamics are consistent with non-interacting, sinusoidal ferromagnetic magnons (Fig. \ref{fig:magnons_vs_temperature}a,c,e,g). Above the Curie temperature, $T_C\approx 305$ K, the ferromagnetic magnons disappear (Fig. \ref{fig:magnons_vs_temperature}b,d,f,h). Across the entire Brillouin zone, the low-temperature magnons in La$_{0.8}$Sr$_{0.2}$MnO$_3$ are nearly resolution-limited and their dispersions are essentially sinusoidal. For quantitative analysis, we modeled the magnon spectra using the simplest applicable approach: a first-nearest-neighbor Heisenberg model \cite{perring1996spin,moudden1997spin} solved within the framework of linear spin wave theory (LSWT). Using the conventions of \textsc{spinw} \cite{toth2015linear}, we obtained a refined exchange constant of $J=8.13$ meV by fitting to experimental data along $(h,0,0)$ (see the Supplementary Information \cite{suppinfo}). With the instrument energy-resolution broadening and momentum integration, magnons from the LSWT calculations in the first-neighbor Heisenberg model agree nearly perfectly with the experimental data (Fig. \ref{fig:main_fig}a). Sometimes fourth-neighbor coupling is included to phenomenologically account for zone-boundary softening \cite{woods2001magnon,furukawa1999magnon,zhang2007magnons,dai2000magnon}; for La$_{0.8}$Sr$_{0.2}$MnO$_3$, including fourth-neighbor coupling does not improve the fit since there is little or no zone-boundary softening. 

Spin-orbit coupling, if present, should induce avoided crossings (gaps) between magnons and phonons. We do not observe any gaps in the magnon dispersion within experimental resolution, which is $\lesssim 2$ meV in the relevant energy range in our $E_i=54$ meV data set (Figs. \ref{fig:main_fig}a and \ref{fig:magnons_vs_temperature}a,c,e,g). At large $|\bm{Q}|$, where the phonon structure factor is strong and the magnon structure factor is weak, there are clear avoided crossings with gaps less than $\approx2$ meV near $\approx16$ meV (cf. Fig. \ref{fig:compare_ins}a). In contrast, at low $|\bm{Q}|$, the phonon structure factor diminishes whereas the magnon contribution dominates. Here (Fig. \ref{fig:magnons_vs_temperature}), magnon dispersions are smooth and continuous, thus any potential avoided crossing must result in a gap that is too small to resolve experimentally. The Supplementary Information includes additional results comparing refined line-cuts to experimental data, highlighting the lack of broadening/softening as well as more detailed analysis of the avoided crossings \cite{suppinfo}.

\begin{figure}
    \centering
    \includegraphics[width=0.5\linewidth]{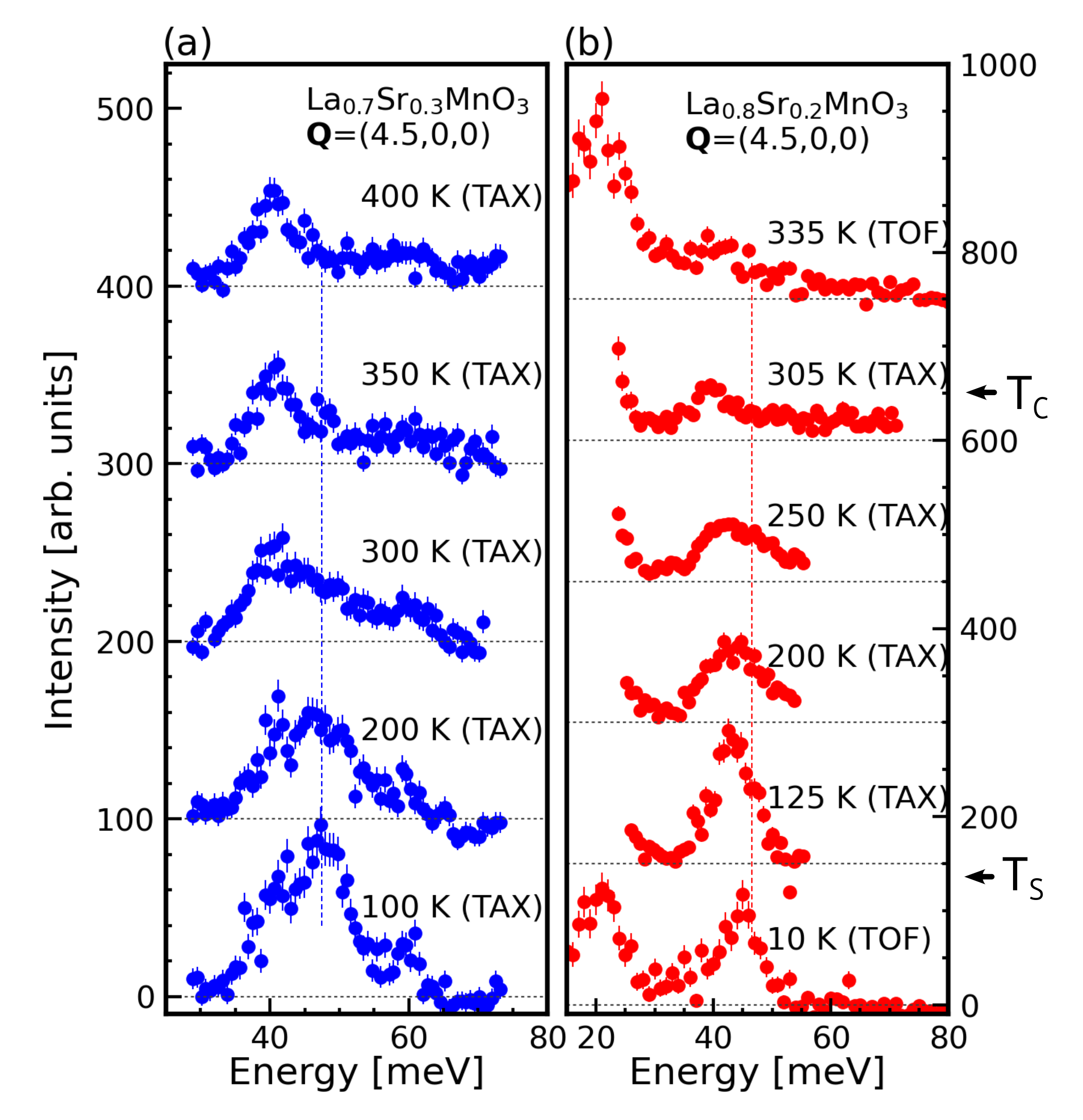}
    \caption{\textbf{Temperature dependence of anomalous phonons in La$_{x}$Sr$_{1-x}$MnO$_3$ ($x=$0.2, 0.3).} (a) TAX neutron scattering intensities from La$_{0.7}$Sr$_{0.3}$MnO$_3$ and (b) TOF and TAX neutron scattering intensities from La$_{0.8}$Sr$_{0.2}$MnO$_3$ at $\bm{Q}=(4.5,0,0)$ as a function of temperature. The data at different temperatures are offset vertically. The structural transition temperature, $T_S\approx120$ K, and the Curie temperature, $T_C\approx305$ K, of La$_{0.8}$Sr$_{0.2}$MnO$_3$ are marked on the right. The thin vertical dashed line goes through the collapsing mode; Multizone fitting shows that the other nearby peak persisting to 335 K is a different phonon.}
    \label{fig:tax_tdep}
\end{figure}

At 335 K ($T_C\approx305$ K), there is pronounced but extremely broad quasielastic magnetic scattering at small crystal momentum in La$_{0.8}$Sr$_{0.2}$MnO$_3$, which looks like conventional paramagnetic spin fluctuations (Fig. \ref{fig:magnons_vs_temperature}). Along the $(h,0,0)$ direction, the magnetic quasielastic scattering is more pronounced than along the $(h,h,0)$ direction, though the cuts along $(h,h,0)$ have larger $|\bm{Q}|$ with weak magnetic signal superposed with phonons.

\subsection{Bond-stretching Mn-O phonons}

\begin{figure}
    \centering
    \includegraphics[width=0.5\linewidth]{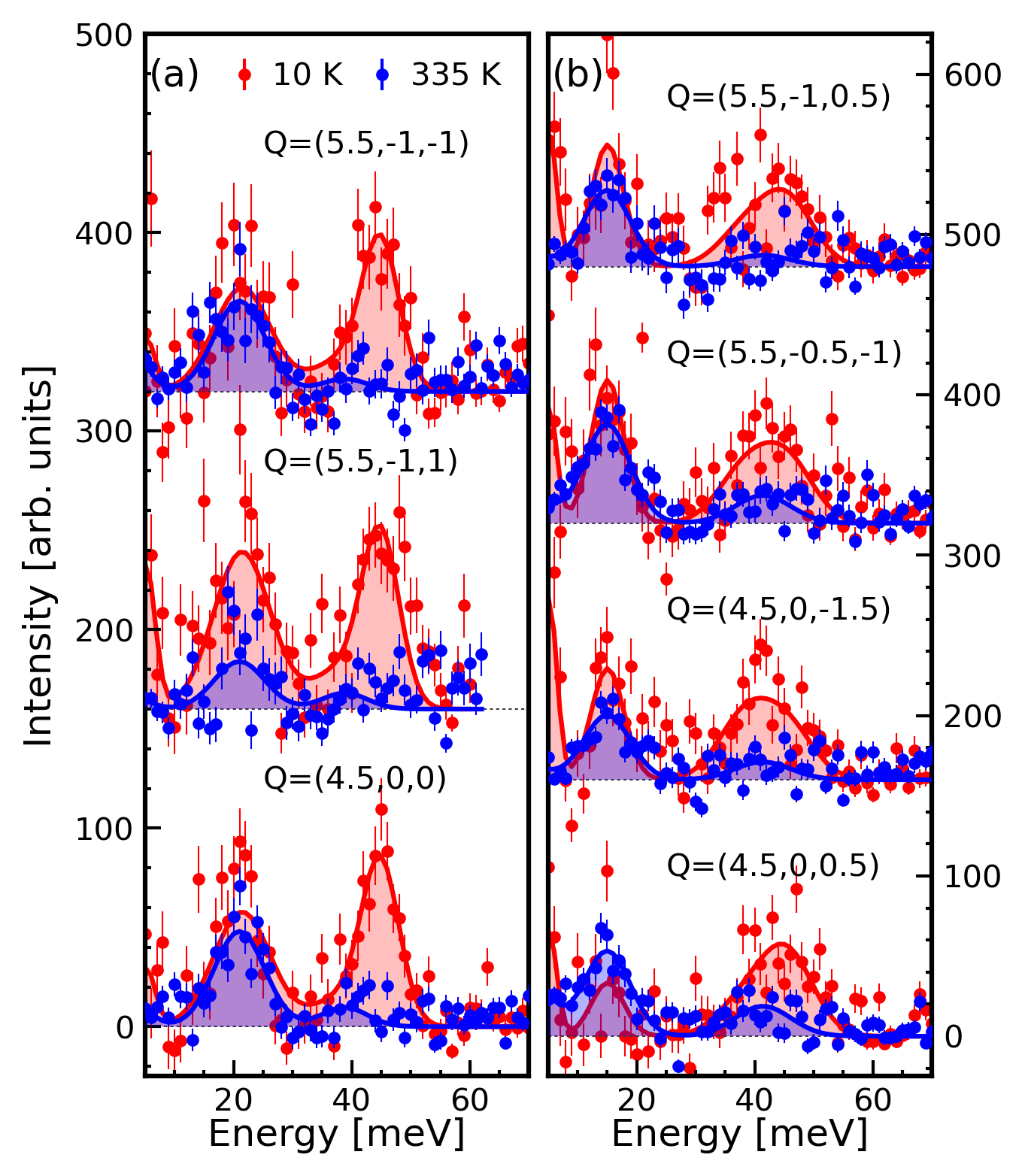}
        \caption{\textbf{Multizone fits of phonons in La$_{0.8}$Sr$_{0.2}$MnO$_3$.} Multizone fits to the inelastic neutron scattering intensities from La$_{0.8}$Sr$_{0.2}$MnO$_3$ at 10 K (red circles) and 335 K (blue circles) with $\bm{q}=(0.5,0,0)$ (a) and $\bm{q}=(0.5,0,0.5)$ (b). The data are from $E_i=120$ meV and were divided by the Bose factor and integrated in a volume $\pm0.1$ rlu. along each cubic axis around each $\bm{Q}$ point.}
    \label{fig:phonon_fits}
\end{figure}

Our inelastic neutron scattering measurements of the phonons revealed anomalous temperature dependence of the Mn-O bond-stretching branch dispersing from $\bm{q}=(0.5,0,0)$ to $\bm{q}=(0.5,0,0.5)$ in Figure \ref{fig:main_fig} (c,d,g) (see also Fig. \ref{fig:compare_ins} where the enhanced intensity at the bottom of the branch is marked by a white circle at $\bm{Q}=(4.5,0,0)$). In La$_{0.8}$Sr$_{0.2}$MnO$_3$, the branch looks conventional at 10 K and is in good agreement with DFT calculations at 0 K (Figs. \ref{fig:main_fig}c,e and \ref{fig:compare_ins}a,c-d). With increasing temperature, the branch gradually broadens and softens (Fig. \ref{fig:tax_tdep}b). The other nearby mode at $\sim 40$ meV is from a non Jahn-Teller phonon (see Fig. \ref{fig:phonon_fits} and the Supplementary Information \cite{suppinfo}). Above the Curie temperature ($T_C\approx 305$ K), the branch is totally collapsed along the entire Brillouin zone edge (cf. Figs. \ref{fig:main_fig}c-d and \ref{fig:phonon_fits}). 

We used the \textsc{phonon explorer} software to perform multizone fitting of the anomalous branch at the $\bm{q} \equiv X = (0.5,0,0)$ and $\bm{q} \equiv M = (0.5,0,0.5)$ points in many Brillouin zones. Consistent with the disappearance of the JT active bond-stretching phonon, the spectra are well described using three peaks at 335 K while at 10 K four are needed for a reasonable fit (Fig. \ref{fig:phonon_fits}). This conclusion is based on data from all zones covered by the data as shown in the Supplementary Information \cite{suppinfo}. Figure \ref{fig:phonon_fits} shows results from selected cuts. We note that we searched for the missing phonon in the entire 335 K dataset containing $\approx$50 Brillouin zones and did not find any detectable intensity that could be attributed to it. Thus we can rule out a structure factor effect causing the scattering intensity to disappear from one zone and appear in others. This observation contrasts with the zone center counterpart of this phonon, which remains strong in several Brillouin zones at 335 K (data not shown).

At 10 K, the zone boundary bond-stretching phonon is centered at $E\approx 45$ meV at $X$ and is well-defined, although already broadened, with Gaussian full width at half maximum (FWHM) $\gamma \approx 7.5 $ meV (Fig. \ref{fig:phonon_fits}a). In the cubic unit cell, this branch corresponds to the planar half-breathing phonon \cite{reznik2005bond} (Fig. \ref{fig:main_fig}g and \ref{fig:phonon_dispersions}e). The underlying distortions of the Mn-O octahedra in the orthorhombic and rhombohedral phases mix the bond-stretching and bond-bending branches \cite{reznik2005bond,reichardt1999anomalous} to a certain extent, but the eigenvectors are similar enough to the typical perovskite bond-stretching phonons that we still discuss them in the conventional language \cite{sterling2021effect}. At the $M$ point, the relevant mode centered at $\approx 45$ meV with FWHM $\approx$ 10 meV corresponds to the quadrupolar bond-stretching phonon (Fig. \ref{fig:phonon_fits}b). At 335 K, the peaks from the half-breathing mode at the $X$ point and the quadrupolar mode at the $M$ point disappear completely, while the peaks from other phonons remain relatively sharp and strong (cf. Figs. \ref{fig:main_fig}c-d,g and \ref{fig:phonon_fits}). 

\begin{figure}
    \centering
    \includegraphics[width=0.5\linewidth]{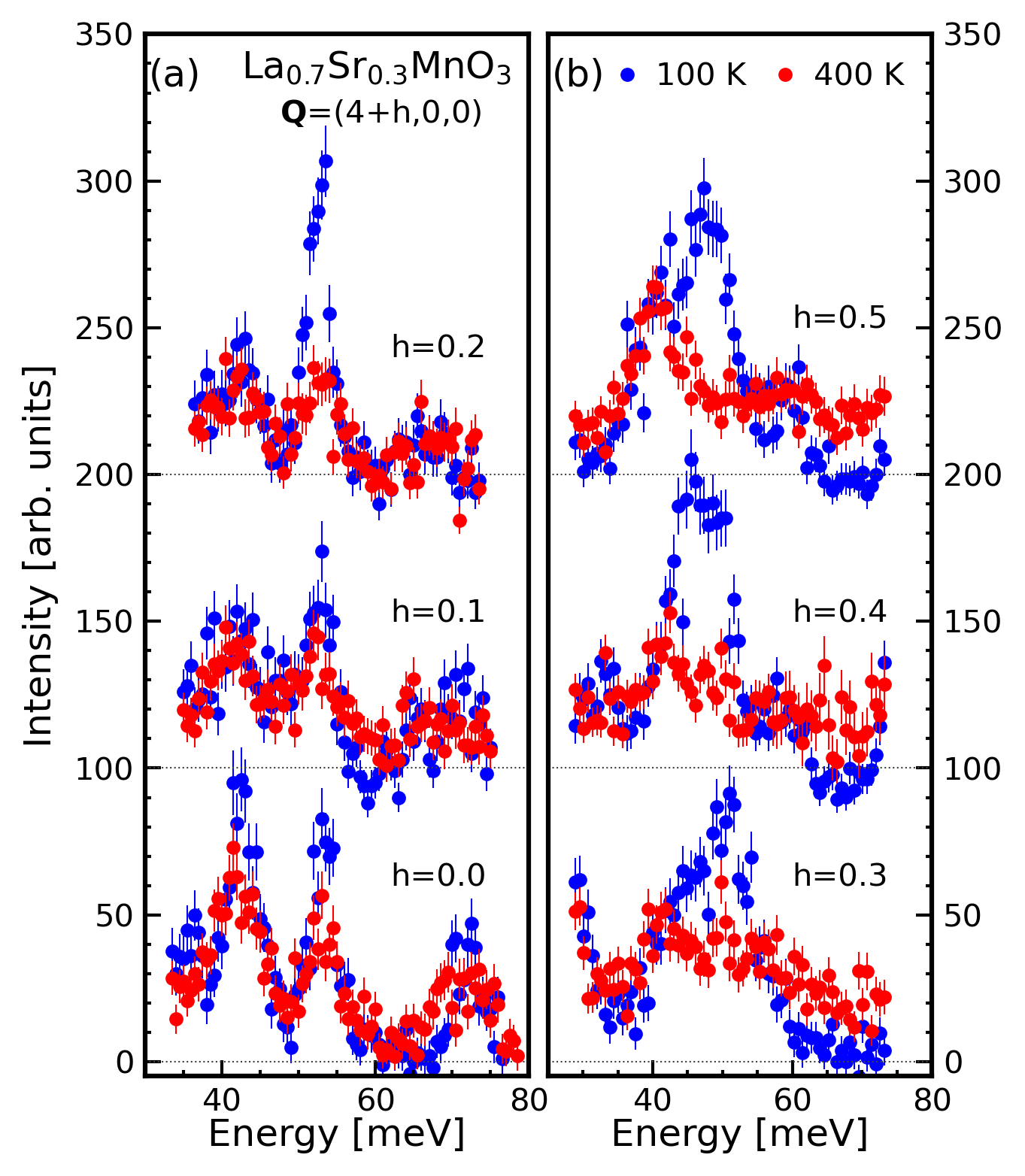}
    \caption{\textbf{$\bm{Q}$ dependence of TAX scattering from La$_{0.7}$Sr$_{0.3}$MnO$_3$.} Triple-axis neutron scattering intensities from La$_{0.7}$Sr$_{0.3}$MnO$_3$ in the $\bm{Q}=(4+h,0,0)$ zone at 100 K (blue circles) and 400 K (red circles). Data for different $h$ are offset vertically. The intensities are divided by the Bose factor.}
    \label{fig:tax_disp}
\end{figure}

Away from the high symmetry zone boundary wave vectors, our TAX data on La$_{0.7}$Sr$_{0.3}$MnO$_3$ are of higher quality than our TOF data on La$_{0.8}$Sr$_{0.2}$MnO$_3$. Figure \ref{fig:tax_disp} illustrates the behavior of the phonon anomaly in La$_{0.7}$Sr$_{0.3}$MnO$_3$ between the zone center and $\bm{q}=(0.5,0,0)$. The behavior at the zone center, characterized by broadening and softening, is conventional for all phonons. The dramatic change of intensity around 45 meV onsets at $h$=0.2 and persists all the way to the zone boundary. It is important to note that the bond-stretching phonon branch gradually increases its Jahn-Teller character on approach to the zone boundary. At the zone boundary, the temperature dependence of the Jahn-Teller mode is similar to that of La$_{0.8}$Sr$_{0.2}$MnO$_3$: at low temperature the Jahn-Teller active branch around 45 meV is strong. Above the Curie temperature, $T_C\approx350$ K, the Jahn-Teller mode disappears leaving behind a weaker Jahn-Teller inactive mode of primarily bond-bending character that is also evident at 10 K.

\begin{figure}
    \centering
    \includegraphics[width=0.5\linewidth]{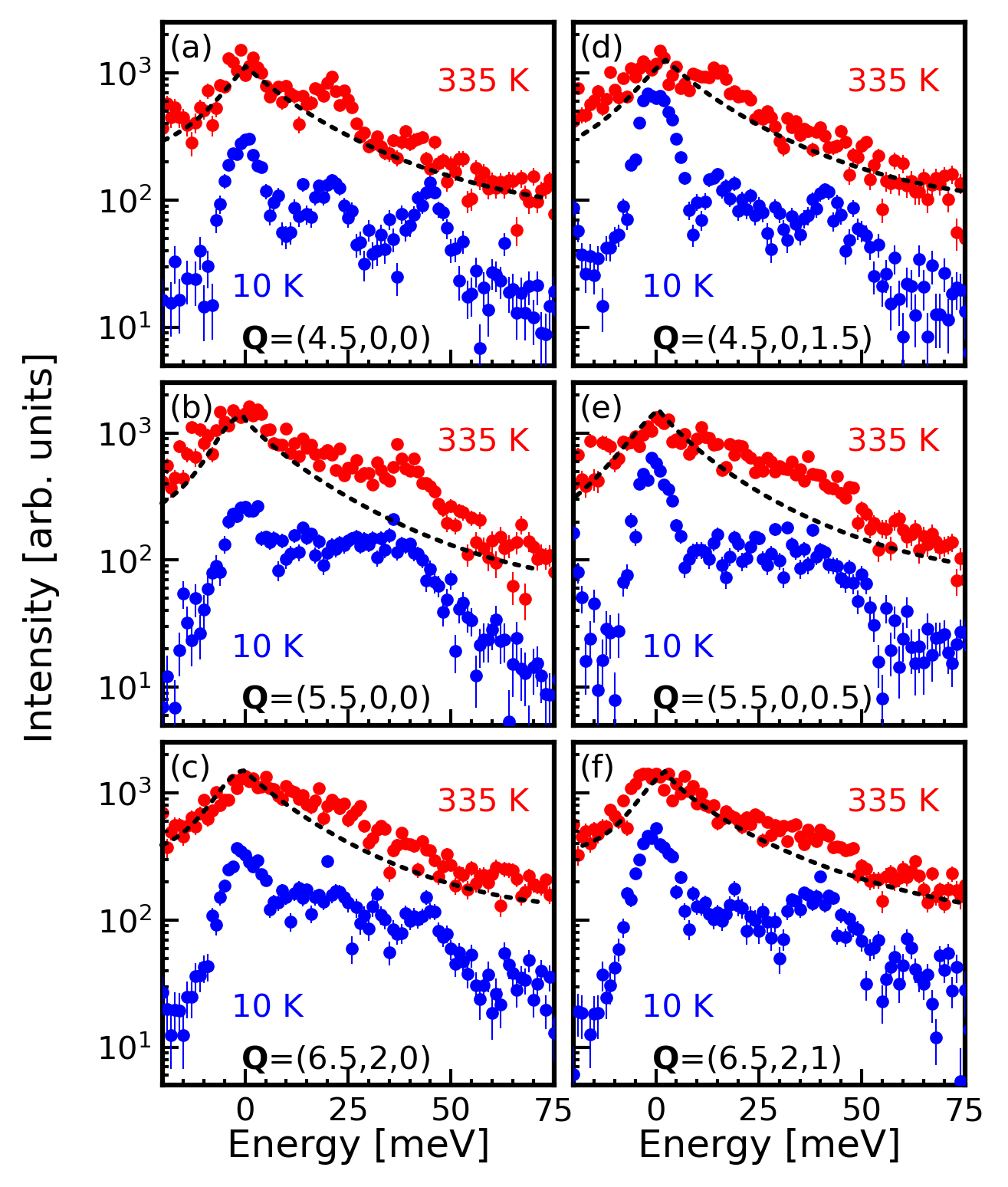}
    \caption{\textbf{Quasielastic neutron scattering from La$_{0.8}$Sr$_{0.2}$MnO$_3$.} Constant momentum cuts of the quasielastic neutron scattering intensities from La$_{0.8}$Sr$_{0.2}$MnO$_3$ at 10 K (blue circles) and 335 K (red circles) with $\bm{q}=(0.5,0,0)$ (a,b,c,f) and $\bm{q}=(0.5,0,0.5)$ (d,e). The data are from $E_i=120$ meV and were integrated $\pm0.1$ rlu. The $\bm{Q}$-points are labeled in each panel. The dashed lines on the data indicate the quasielastic background; they are a guide to the eye. The data at 335 K are offset vertically for clarity (scaled by 4 in linear scale).}
    \label{fig:quasi_elastic_linecuts}
\end{figure}

\subsection{Quasielastic scattering in La$_{0.8}$Sr$_{0.2}$MnO$_3$}

A sum rule ensures that the intensity integrated over all energies is conserved, so the spectral weight of the collapsed modes redistributes elsewhere in reciprocal space at high temperatures. Our TOF data from La$_{0.8}$Sr$_{0.2}$MnO$_3$ show that the peaks at lower energy neither broaden substantially nor gain intensity (Fig. \ref{fig:phonon_fits}), so the spectral weight from the collapsed mode does not redistribute into \textit{finite} lower energy phonon peaks. We argue that at least some of this spectral weight is transferred into quasielastic scattering. Earlier work identifies quasielastic intensity underneath the $E=0$ signal in proximity to strongly damped transverse acoustic modes \cite{maschek2016polaronic,maschek2018polaronic,helton2019damping}. We find strong quasielastic scattering at several large zone boundary wavevectors (Figs. \ref{fig:quasi_elastic_linecuts} and \ref{fig:main_fig}d) showing a broad central peak at 335 K. These  are far from the wave vectors investigated before \cite{maschek2016polaronic,maschek2018polaronic,helton2019damping}. No physically meaningful $\bm{q}$-space structure in elastic or low energy inelastic diffuse scattering is evident from the data. Large wave vector of this signal rules out any magnetic origin.

\section{Discussion} 

Previous neutron scattering studies in the low $T_C$, high CMR manganites have found an assortment of phonon anomalies characteristic of the expected giant EPC \cite{fernandez2006spin,shimomura1999x,adams2000charge,dai2000short,zhang2001jahn}. To be specific, we take giant EPC to mean a non-perturbative renormalization of the phonons, like total collapse of a mode, as contrasted to perturbative renormalization like energy shifts and broadening. Here we found such non-perturbative renormalization of the JT active bond-stretching optical phonons in both La$_{0.7}$Sr$_{0.3}$MnO$_3$ and La$_{0.8}$Sr$_{0.2}$MnO$_3$. 

To be sure the anomalous phonon effect is due to EPC, we need to rule out both spin-phonon coupling and the orthorhombic to rhombohedral structural phase transition at $\approx 120$ K in La$_{0.8}$Sr$_{0.2}$MnO$_3$ \cite{paul1998influence,dabrowski1999structure,kamenev1997influence,asamitsu1996magnetostructural,tokura1994giant,urushibara1995insulator} accompanied by twinning as the origin of the effect. We performed detailed temperature-dependence measurements and DFT calculations for both phases. A strong zone boundary phonon peak predicted by both calculations in the vicinity of 45 meV is experimentally observed at low temperatures (Fig. \ref{fig:main_fig}e-f). Calculations show that eigenvectors of this mode have mostly half-breathing bond-stretching character at the $X$ point and quadrupolar, in-plane bond-stretching character at the $M$ point (cf. Fig. \ref{fig:phonon_dispersions}e). The differences in eigenvectors from the typical cubic perovskite phonons are due to the reduced symmetry in the orthorhombic and rhombohedral phases, which mixes the bond-stretching modes with other branches in the folded Brillouin zone \cite{reznik2005bond}. The good agreement between low temperature data and the calculations for both phases suggests that the anomalous JT active branch is insensitive to the structural phase transition. In fact TAX measurements of the JT active mode at $\bm{Q}=(4.5,0,0)$ show that the bond-stretching branch does not change much across the structural phase transition temperature of 120 K, but becomes colossally damped around the Curie temperature of $T_C\approx 305$ K (Fig. \ref{fig:tax_tdep}). We further ruled out twinning by calculating the neutron scattering intensity for all permutations of lattice vectors in both phases; all orientations were in good agreement with the experimental data, suggesting that the anomalous JT active branch is insensitive to twinning. A good agreement between DFT in either phase and the TOF data suggests that the lattice dynamics of La$_{0.8}$Sr$_{0.2}$MnO$_3$ are conventional at low temperature.

In the magnetic channel, low temperature La$_{0.8}$Sr$_{0.2}$MnO$_3$ behaves like a conventional Heisenberg ferromagnet with sharp, sinusoidal magnon dispersions \cite{perring1996spin,moudden1997spin,furukawa1996spin} and no evidence for magnon-phonon hybridization \cite{hennion2019interacting,petit2009quantized} (Figs. \ref{fig:magnons_vs_temperature} and the Supplementary Information \cite{suppinfo}) despite strong EPC in other channels (e.g. charge or orbital). This result is expected from small spin-orbit coupling in the manganites, which controls the magnitude of magnon-phonon hybridization. In fact we expect no such hybridization in any materials with weak spin-orbit interaction including manganites at lower doping levels. Moreover, we don't observe any EPC induced renormalization or any other clearly identifiable spin-phonon interaction in La$_{0.8}$Sr$_{0.2}$MnO$_3$ \cite{dai2000magnon,zhang2007magnons,ye2006evolution,ye2007spin}. This behavior is different from earlier reports on lower T$_C$ manganites that reported significant broadening and magnon branch flattening \cite{dai2000magnon}.

Based on the agreement of the neutron scattering with DFT phonons and nearest-neighbor Heisenberg model magnons (Fig. \ref{fig:main_fig}), the low temperature ferromagnetic phase has conventional atomic lattice and magnetic dynamics. However, optical phonons of JT character expected from DFT disappear from the inelastic spectrum above T$_C$. 

Most likely the JT branches collapse into quasistatic, charge-trapping distortions that diffuse throughout the lattice, contributing to the current in CMR; neutron scattering from the quasistatic distortions produces the quasielastic signal centered at zero energy in Figure \ref{fig:quasi_elastic_linecuts}. The absence of momentum space structure in this quasielastic signal above $T_C$ indicates incoherent, distortions lacking any well-defined short or long range order. It has been suggested that the magnetic quasielastic scattering at low $|\bm{Q}|$ above $T_C$ (Fig. \ref{fig:magnons_vs_temperature}) could originate from fluctuating, inhomogeneous ferromagnetic domains pinned to the quasitatic JT disortions \cite{adams2000charge,zhang2001jahn}; this picture is suggestive of Griffith's effects in manganites \cite{griffiths1969nonanalytic,jiang2007griffiths,jiang2007extreme,jiang2008griffiths,saha2022origin,krivoruchko2010griffiths,salamon2003griffiths,ghorai2021effect}.

The bond-stretching branch disappears across an entire Brillouin zone edge rather than at just a single point.
This is significant because polaronic hopping underlying transport in CMR can not be described by a purely local JT distortion (c.f. Holstein phonons) since hopping from one site to another depends on the distortion at each site \cite{maschek2016polaronic,maschek2018polaronic}. Such a process requires extended phonons (i.e. Bloch functions) that connect distortions between neighboring Mn sites. Our work directly identifies the implicated phonons that participate in CMR and elucidates the dynamic state above $T_C$. If we imagine that a cooperative JT distortion ``freezes in" bond-stretching phonons with random phases, then neighboring Mn-O octahedra are necessarily coupled with a net distortion that looks like an incoherent superposition of half-breathing, full-breathing, and quadrupolar modes. We cannot resolve the full-breathing phonon in our inelastic scattering data due to kinematic constraints, but we find that the half-breathing and quadrupolar phonons are indeed entirely collapsed above $T_C$, corroborating this picture.

The prevailing theory of CMR correlates EPC strength and the magnitude of CMR. Then low CMR compounds like La$_{1-x}$Sr$_x$MnO$_3$ ought to show weak EPC effects compared to high CMR compounds; however, they show non-perturbative atomic lattice distortions. This is a signature of giant EPC, even when the high temperature paramagnetic phase is metallic \cite{louca1997local,louca1999local,maschek2016polaronic,maschek2018polaronic}. Furthermore, diffraction experiments \cite{weber2013large} show that the oxygen atom displacements due to lattice distortions are comparable in magnitude between weak CMR material La$_{0.7}$Sr$_{0.3}$MnO$_3$ and compounds with strong CMR like La$_{0.65}$Ca$_{0.35}$MnO$_3$ and La$_{1.2}$Sr$_{1.8}$Mn$_2$O$_7$ \cite{maschek2016polaronic,maschek2018polaronic,dai1996experimental}. Inelastic scattering from damped acoustic phonons also shows that EPC is similarly strong in La$_{0.8}$Sr$_{0.2}$MnO$_3$ and La$_{0.7}$Sr$_{0.3}$MnO$_3$. This is all despite an order of magnitude stronger CMR in La$_{0.8}$Sr$_{0.2}$MnO$_3$ than in La$_{0.7}$Sr$_{0.3}$MnO$_3$ \cite{maschek2018polaronic}. In apparent support of the theory correlating CMR with EPC, the collapse of JT phonons has been observed before in high CMR La$_{0.7}$Ca$_{0.3}$MnO$_3$ \cite{zhang2001jahn}. However, the unexpected collapse of the same modes in comparatively weak CMR La$_{1-x}$Sr$_x$MnO$_3$ compounds violates this relationship. We not only observe the collapse of the Mn-O-Mn bond-stretching phonon with JT character along $(h,0,0)$, as reported before \cite{zhang2001jahn}, but also along the entire zone boundary from $(1/2,0,0) \rightarrow (1/2,1/2,0)$ in La$_{0.8}$Sr$_{0.2}$MnO$_3$ (Fig. \ref{fig:main_fig}c,d). In short, there is evidence for strong EPC across the whole spectrum of CMR magnitudes. 

These findings cast doubt on any simple relationship between EPC strength and the magnitude of CMR. Our work demonstrates that Jahn-Teller-active optical phonons, believed to be at the heart of CMR, also show very strong renormalization and total collapse in FM manganites which have small CMR. The crucial difference between low CMR and high CMR compounds appears to be not the strength of EPC or magnitude of lattice distortions in the paramagnetic phase but diffusive dynamics of these distortions acting as the lattice component of current-carrying polarons. They are slow, in fact often static, in the high CMR compounds and relatively fast in the low CMR compounds \cite{maschek2016polaronic,maschek2018polaronic}. These results necessitate development of new quantitative models of the relationship between structure/chemistry and electronic transport in FM manganites that focus on distortion diffusion rates rather than on distortion amplitudes.

\section{Conclusion}

We measured the phonon spectra of high T$_C$ La$_{0.7}$Sr$_{0.3}$MnO$_3$ using triple-axis neutron scattering and both the magnon and phonon spectra of high $T_C$ La$_{0.8}$Sr$_{0.2}$MnO$_3$ across a broad range of momentum and energy using inelastic time-of-flight neutron scattering and found a total collapse of JT active bond-stretching phonons as temperature increased above the Curie temperature. DFT and LSWT calculations combined with neutron scattering data ruled out structural phase transitions, twinning, and spin-phonon interactions as mechanisms for this effect. Magnon renormalization that could be attributed to spin-phonon coupling was negligible in La$_{0.8}$Sr$_{0.2}$MnO$_3$, so the phonon renormalization is likely driven by giant EPC due to JT coupling. Since La$_{0.7}$Sr$_{0.3}$MnO$_3$ and La$_{0.8}$Sr$_{0.2}$MnO$_3$ both have weak CMR but giant renormalization of the JT phonons, we conclude that strong EPC of JT type is not sufficient for strong CMR. Our work directly identifies optical phonons that dominate in EPC in La$_{0.8}$Sr$_{0.2}$MnO$_3$ and provides strong evidence that the accepted picture connecting EPC, CMR, and spin-phonon interactions in manganites needs to be reexamined.

\section{Methods}

\subsection{Neutron scattering experiments}

Time-of-flight (TOF) neutron scattering measurements were performed on La$_{0.8}$Sr$_{0.2}$MnO$_3$ on the 4SEASONS chopper spectrometer at the Materials and Life Science Experimental Facility at J-PARC \cite{kajimoto2011_4seasons}. This spectrometer allows simultaneous use of multiple incident neutron energies, $E_i$, providing datasets with different resolutions and energy/momentum ranges \cite{nakamura2009multiei}. We obtained data with $E_i=120$ meV providing elastic resolution of 8 meV and $E_i = 54$ meV providing elastic resolution of 3 meV by rotating the Fermi chopper at 300 Hz (see the Supplementary Information for energy-dependent resolution \cite{suppinfo}). The sample was mounted with the [100]/[010] scattering plane horizontal and was rotated during the experiment around the vertical axis over 100$^{\circ}$ (at 10 K) and 40$^{\circ}$ (at 335 K). This configuration covered a large number of Brillouin zones of interest, with the exception of zones providing access to longitudinal phonons dispersing along $(110)$. The TOF data were post processed with the \textsc{phonon explorer} software \cite{reznik2020automating} which leverages large reciprocal space coverage to fit data in many Brillouin zones simultaneously using the multizone fitting scheme \cite{parshall2014phonon}. \textsc{phonon explorer} uses \textsc{mantid} to reduce and histogram data \cite{ARNOLD2014156,savici2022efficient}.

Triple-axis (TAX) experiments were carried out on both La$_{0.8}$Sr$_{0.2}$MnO$_3$ and La$_{0.7}$Sr$_{0.3}$MnO$_3$ at the 1T spectrometer located at the ORPHEE reactor. The experimental conditions were the same as in ref. \cite{reznik2005bond}. 

La$_{1-x}$Sr$_{x}$MnO$_3$ are perovskites with orthorhombic/rhombohedral structural distortions at low temperature \cite{paul1998influence,dabrowski1999structure,kamenev1997influence,asamitsu1996magnetostructural,tokura1994giant,urushibara1995insulator}. The distortions are small enough that it is convenient to study the lattice dynamics in the pseudocubic basis \cite{reznik2005bond,weber2013large,maschek2016polaronic,maschek2018polaronic}. We use pseudocubic notation with $a\approx3.9~\textrm{\AA}$ throughout. The conversion between the orthorhombic/rhombohedral and pseudocubic bases is discussed in the Supplementary Information \cite{suppinfo}. The structural distortions affect the bond-stretching phonons by mixing them with other branches \cite{reznik2005bond}, but they retain their bond-stretching character (see below). 

\subsection{Density functional theory calculations}

Phonon eigenvectors and inelastic neutron scattering intensities from phonons in La$_{0.8}$Sr$_{0.2}$MnO$_3$ were calculated using density functional theory (DFT) in the \textsc{abinit} package \cite{gonze2020abinit,Romero2020} using the PBE functional \cite{perdew1996generalized}. We used the projector augmented wave (PAW) method \cite{Torrent2008} with a 500 eV energy cutoff in the interstitial region and a 750 eV cutoff in the atomic spheres. PAW datasets were retrieved from \textsc{pseudodojo} \cite{jollet2014generation,van2018pseudodojo}. To remedy the well known delocalization error in DFT, we included a 4 eV Hubbard-U potential on the Mn $d$ orbitals \cite{Amadon2008}. We simulated Sr doping in La$_{0.8}$Sr$_{0.2}$MnO$_3$ by explicitly removing 0.2 electrons from the simulation cell for each formula unit and adding a neutralizing positive charge background (we call this ``charge doping") \cite{bruneval2014consistent}. Besides neglecting disorder induced inhomogeneity broadening \cite{roy2023occupational}, we postulate that charge doping instead of explicit Sr substitution is sufficiently accurate. See the Supplementary Information for a more detailed discussion \cite{suppinfo}. All calculations were done in the ferromagnetic phase.

In the orthorhombic phase, we used an $8 \times 6 \times 8$ $\bm{k}$-point grid. We used \textsc{phonopy} to calculate force constants in a $2 \times 1 \times 2$ orthorhombic supercell \cite{phonopy-phono3py-JPCM,phonopy-phono3py-JPSJ}. The relaxed, charge doped orthorhombic structure is predicted to be stable, i.e. there are no imaginary phonons. In the rhombohedral phase, we used an $8 \times 8 \times 8$ $\bm{k}$-point grid during relaxation and a $2 \times 2 \times 2$ supercell for phonons. The rhombohedral phase is unstable at 0 K, as expected, with the instability coming from a low energy optical branch corresponding to octahedral rotations. The unstable mode has minimal effect on the relevant bond-stretching phonons since they are non-degenerate normal modes. 

We used \textsc{euphonic} to calculate inelastic neutron scattering intensities from the phonon eigenvectors \cite{fair2022euphonic}. To approximate experimental resolution, we used a Gaussian energy broadening with FWHM=5 meV (accurate at $E\approx45$ meV with $E_i=120$ meV, see the Supplementary Information for more details \cite{suppinfo}). To present the calculated intensities in the cubic notation, we ``unfolded" by calculating the intensities in the orthorhombic and rhombohedral cells using wave vectors in the r.l.u. of the pseudocubic basis \cite{sterling2024lattice,sterling2021effect}. Further details of the DFT calculations and discussion of results not shown below are relegated to the Supplementary Information \cite{suppinfo}.

\subsection{Spin wave calculations}

Magnon spectra and neutron scattering from magnons in La$_{0.8}$Sr$_{0.2}$MnO$_3$ were calculated in linear spin wave theory (LSWT) using \textsc{spinw} \cite{toth2015linear}. Ferromagnetic exchange, $J$, was extracted by first ``coarse" fitting multiple random slices from the 10 K experimental data from both incident energies simultaneously. The coarse fit results are shown in the Supplementary Information \cite{suppinfo}; the result was $J=8.52$ meV in the \textsc{spinw} conventions. Starting from the coarse fit, the exchange parameter was refined by fitting to a small selection of experimental cuts along $(h,0,0)$ from $E_i=54$ meV. For the refined fit, we used the actual energy-dependent energy resolution and broadened over momentum to account for momentum resolution from e.g. crystal mosaic (see the Supplementary Information for details of momentum resolution \cite{suppinfo}). Since the magnon dispersions are very steep, converging the model momentum resolution required integration over a very dense Gaussian weighted $30\times30\times30$ grid of $\bm{q}$-points of interest. Thus, we only fit a small subset of $\bm{q}$-points in the refinement of $J$; the result after refinement was $J=8.13$ meV. The magnon dispersion with $J=8.13$ meV is shown in Figure \ref{fig:main_fig}a (also see the Supplementary Information for more results of the refined fits \cite{suppinfo}).

\backmatter

\bmhead{Acknowledgments}

T. C. S. and D. R. were supported by the U.S. Department of Energy, Office of Basic Energy Sciences, Office of Science, under Contract No. DE-SC0024117. N. K. was funded by the Deutsche Forschungsgemeinschaft (DFG, German Research Foundation) under Project No. 419331252. The neutron scattering experiments at J-PARC were performed under a user program (No. 2020B0160). This work used the Alpine high performance computing resource at the University of Colorado Boulder \cite{alpine}. Alpine is jointly funded by the University of Colorado Boulder, the University of Colorado Anschutz, Colorado State University, and the National Science Foundation (award 2201538).

\bmhead{Author contributions}

T.C.S. and D.R. wrote the manuscript. T.C.S. and D.R. analyzed the data.  T.C.S. performed the calculations. D.R. and F.W. conceived and supervised the project. R.K. and K.I. did the TOF neutron scattering experiment. N.K. and A.T.S assisted with handling/reducing the TOF data. D.R. and F.W. did the TAX neutron scattering experiments. All authors reviewed and edited the text.

\bmhead{Supplementary information}

The supplementary information contains additional details of the density functional theory calculations and more comparison to experimental neutron scattering. Also included are the results of multizone refinement fitting of the neutron scattering intensity.

\bmhead{Data Availability}

All data are available from the corresponding authors upon reasonable request.

\bmhead{Competing Interests}

The authors declare no competing interests.



\end{document}